# Multi-User Activity Recognition and Tracking Using Commodity WiFi

**Sheng Tan, Linghan Zhang, Zi Wang, Jie Yang**
Florida State University, Tallahassee, Florida
{tan,lzhang,ziwang,jie.yang}@cs.fsu.edu

## ABSTRACT

This paper presents MultiTrack, a commodity WiFi based human sensing system that can track multiple users and recognize activities of multiple users performing them simultaneously. Such a system can enable easy and large-scale deployment for multi-user tracking and sensing without the need for additional sensors through the use of existing WiFi devices (e.g., desktops, laptops and smart appliances). The basic idea is to identify and extract the signal reflection corresponding to each individual user with the help of multiple WiFi links and all the available WiFi channels at 5GHz. Given the extracted signal reflection of each user, MultiTrack examines the path of the reflected signals at multiple links to simultaneously track multiple users. It further reconstructs the signal profile of each user as if only a single user has performed activity in the environment to facilitate multi-user activity recognition. We evaluate MultiTrack in different multipath environments with up to 4 users for multi-user tracking and up to 3 users for activity recognition. Experimental results show that our system can achieve decimeter localization accuracy and over 92% activity recognition accuracy under multi-user scenarios.

## 1 INTRODUCTION

Indoor human tracking and activity recognition is gaining increasing attention and undergoing fast development in a variety of real-world applications, especially in human-computer interaction (HCI) area. Particularly, indoor human tracking is a building block for more comprehensive context-based services that enable interaction between cyberworld and physical world. For example, it provides human computer interface for visually impaired people to explore and navigate surrounding areas and receive location-based services [15]. In addition, human activity recognition can be naturally intergraded with a broad array of applications that require cyber-physical interactions, such as in smart home, virtual/augmented reality, gaming and exercise monitoring [3, 23, 25, 30, 32]. Tracking human location and activity can also be used to monitor well-being and suggest behavioral changes for people with special needs [7, 15, 34]. Existing work in indoor tracking and activity recognition mainly relies on dedicated sensors (i.e., RFID, motion sensors, mobile device) [4, 36, 43, 49, 51] that are worn/carried by the user or depth/infrared cameras and visible light sensors [6, 10, 12] that are installed in the environment (e.g., Kinect, leap motion, light sensor [17, 18, 22]). These solutions require significant deployment overhead and incur non-negligible cost. Moreover, the camera and visible light based approaches cannot work in non-line-of-sight (NLOS) scenarios and often involve user privacy concerns. The systems that rely on sensors worn/carried by the user could be inconvenient and cumbersome as they require user's explicit involvement. For example, users at home especially elderly and children may forget to carry the device or might be reluctant or feel uncomfortable to carry tracking devices. Recently, Radio Frequency (RF) based device-free human sensing becomes an appealing alternative. It analyzes the radio signal reflections from human body for human tracking and activity recognition thus doesn't require user to wear or carry any sensor. It also provides better coverage and works under NLOS scenario as wireless signals can penetrate walls when compared to camera or visible light based approach. Existing work in RF based device-free sensing uses either specialized (e.g., USRP) [1, 2, 27] or commodity hardware [29, 35, 45]. In this work, we focus on latter approach as it can reuse existing WiFi infrastructure to facilitate easy and large-scale deployment without incurring additional cost due to the proliferation of WiFi devices and networks.

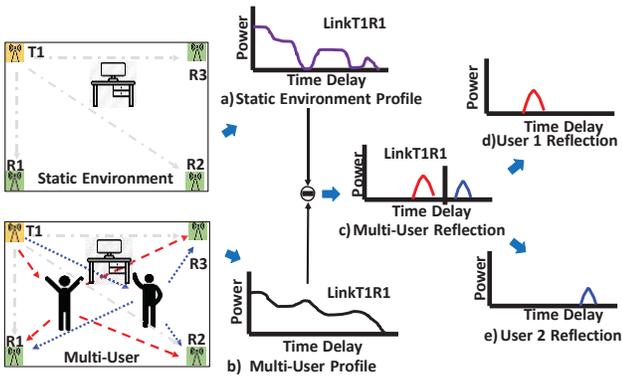

Figure 1: Basic idea of MultiTrack.

However, current commodity WiFi sensing systems are mainly designed for and tested with the presence of a single user in the physical environment. They cannot work well when multiple users are present in the same environment simultaneously. This is because the received signals are mixture of signal reflections from different users and these systems cannot identify the signal reflections that correspond to each individual. WiMU [38] attempts to address this issue by searching the possible combination of multiple known gestures based on the assumption that all gestures in each set of simultaneously performed gestures are predefined. However, such an assumption often cannot hold in practice for multiple user tracking as the walking trajectory of each individual tends to be random and cannot be predefined. Similarly, if one or more users perform unknown activities to WiMU, while other users perform predefined activities, WiMU cannot recognize the predefined activities due to the interference from the users performing unknown background activities [38].

In this paper, we propose MultiTrack, a commodity WiFi based sensing system for multi-user tracking and activity recognition. MultiTrack is able to track the locations of multiple users that walk simultaneously with decimeter accuracy, which is comparable to the accuracy of existing systems that focus on only a single user. In addition, it can recognize the activities from multiple users simultaneously or the activity of the target users when there are other users performing unknown background activities.

The basic idea of our system is to identify and extract the signal reflection from each individual user with the help of multiple WiFi links and all the available WiFi channels at 5GHz. As shown in Figure 1, we can leverage existing WiFi access points and WiFi devices (e.g., desktops, laptops, smart appliances) to form multiple WiFi links. Such WiFi links can quantify the radio signal propagation in terms of the power delay profile, which describes the power intensity of received signals as a function of propagation delay. Figure 1 (a) shows the power delay profile under static environment at the wireless link $LinkT1R1$ without any human presence. Once there are multiple users performing activities or walking simultaneously, we can obtain another power delay profile under multi-user case, as shown in Figure 1 (b). By subtracting the profile under multi-user from the one under static environment, we can obtain the profile of signal reflections that resulted only from the activities of multiple users, as shown in Figure 1 (c). Then, we can segment the multi-user reflection profile into single user reflection profile, which corresponds to the signal reflected from each individual user. By analyzing the single user reflection profile at multiple links, we are able to perform multi-user tracking. Moreover, we can reconstruct the signal profile of each user as if only a single user has performed activity in the environment to facilitate multi-user activity recognition.

Intuitively, we can derive the power delay profile from the Channel State Information (CSI) that measured at each received WiFi packet. However, commonly used WiFi channel that used to send out each WiFi packet only has 20MHz/40MHz bandwidth, which provide a time or distance resolution at 50/25ns or 15/7.5 meters for distinguishing different signal propagations. Such a resolution is larger than the dimension of a typical room and is unable to distinguish signal reflections from different users in confined indoor spaces. Inspired by pioneer work on channel splice [48], we propose to send out probe signals at all available channels of 5GHz (i.e., over 600MHz). We then combine all the channels at 5GHz to derive a fine-grained power delay profile, which is used to separate the signal reflection from each individual user.

Moreover, even with stitched channels that spread over 600MHz bandwidth at 5GHz frequency, the reflected signal from users within close proximity could still partially mix together. And such a scenario could be quite common when the indoor environment has limited space (i.e., small office, bedroom). To solve this problem, instead of modeling each human as a single reflector in existing work [19, 28], we propose to model each human into primary reflector (i.e., upper body) and secondary reflector (i.e., arms, legs and head). By incorporating primary and secondary reflector model with fine-grained power delay profile, we are able to obtain a more accurate user reflection profile for user tracking. For activity recognition, our system extracts Doppler frequency shift based feature from the signal reflection of each user, which isolates the signal dynamic due to human activity from the signal reflected from static objects and walls. The extracted feature is then compared against the features of known activities to facilitate multi-user activity recognition.

We experimentally evaluate MultiTrack in three different indoor environments (i.e., home, classroom and corridor) with up to four users that are walking simultaneously. For multi-user activity recognition, we evaluate our system using six different body weight exercises (sit up, squat, lunge, spinal balance, bicycle crunch, and toe-touch crunch) with up to three users performing either pre-defined or unknown activities simultaneously. The results shows MultiTrack achieves high recognition accuracy even when non-target users are performing unknown background activities. The contribution of our work are summarized as follows:

- We show that the commodity WiFi can be utilized to perform multi-user tracking and activity recognition. Such an approach does not require any dedicated or specialized devices and can work under NLOS scenarios.
- We leverage the large bandwidth at 5GHz to extract fine-grained power delay profile at multiple WiFi links to disentangle signal reflection from multiple users in the multipath rich environment. We model the human body as primary and secondary reflectors to further separate users within close proximity for improved tracking accuracy.
- We conduct extensive experiments in different multipath environments. Experimental results show that MultiTrack achieves decimeter localization accuracy and over 92% recognition accuracy even when non-target user is performing unknown activities simultaneously in the background.

## 2 SYSTEM DESIGN

### System Overview

The basic idea of our system is to identify and separate the signal reflections from different users by leveraging multiple WiFi links and the large WiFi bandwidth at 5GHz. By analyzing the separated signal reflection of each individual, our system can achieve multi-user tracking and activity recognition. Figure 2 shows the flow of our system. The system first performs channel scanning and CSI collection, in which one WiFi transmitter continuously sends out probe packets through all available channels at 5GHz in each time frame and three or more WiFi receivers extract Channel State Information (CSI) measurement from each received packet. Note that the time is divided into non-overlap time frames and each time frame is less than the coherence time where the multipath environment is considered consistent. The extracted CSI measurements then go through calibration process to mitigate both amplitude and phase errors.

After CSI measurements calibration, our system stitches all the available channels together to derive a fine-grained power delay profile. We adopt inverse non-uniform Discrete Fourier Transform (NDFT) to overcome the problem of unequal and non-contiguity spreading of available channels on 5GHz frequency band that used for commodity WiFi. The derived fine-grained power delay profile at each link then goes through Reflection Profile Segmentation to determine the number of users and to segment the power delay profile into single user reflection profiles, where each one represents the signal reflection dominated by one individual user.

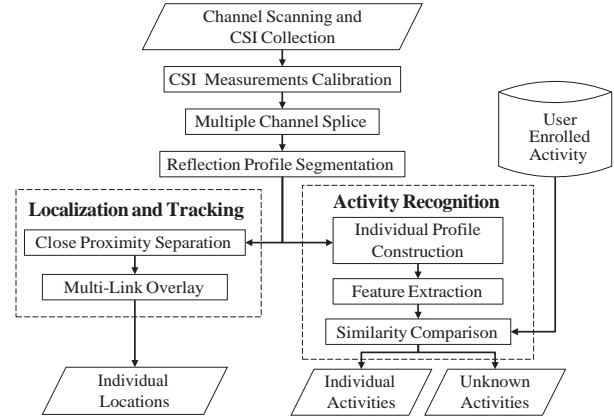

**Figure 2: Overview of system flow.**

Next, our system splits into two subsystems. The first one is for multi-user tracking and the other one is for multi-user activity recognition. For localization and tracking subsystem, we first leverage Close Proximity Separation component, where the human body is modeled as primary and secondary reflectors, to further refine the separated single user reflection profile. Then, we overlay the refined signal reflection profile from multiple links and identify the converged one with highest power as users' locations.

For activity recognition subsystem, we first reconstruct the power delay profile of each user as if only a single user has performed activity in the environment. Then, we extract Doppler frequency shift based on the reconstructed signal profile to isolate the signal reflections from surrounding objects and environments. We next extract energy-based frequency contour in the Doppler frequency shift as feature and calculate the similarity of such feature with respect to each enrolled activity using Muti-Dimensional Dynamic Time Warping (MD-DTW). The one that has the highest and also sufficient similarity compare to the profile in the library is then identified as the recognized activity.

### CSI Collection and Calibration

For system with 802.11n/ac wireless network, we are able to extract channel state information (CSI) from the WiFi NIC. Such CSI can be viewed as a sampled version of the channel frequency response $h(f)$. Particularly, the standard 20/40MHz WiFi channel measures the amplitude and phase information for each of the 56/128 orthogonal frequency-division multiplexing (OFDM) subcarriers. In our work, we utilize the total 24 available 20MHz channels at 5GHz band.

By setting the channel hopping delay as 0.2ms, we are able to ensure our system can hop through all available channels within the coherence time.

Due to hardware limitation of COTS WiFi NICs, the extracted raw CSI measurements involve significant distortions. Such distortions or errors are mainly caused by clock unsynchronization. We adopt the error correction approach from previous work [48] for data calibration. In particular, we mitigate the amplitude error by averaging raw CSI measurements from multiple packets that collected within coherence time. We mitigate the constant phase error by picking a reference channel from all existing channels and compensate the phase difference between each channel pair. The linear component of phase error can be further separated into two parts. The first part of linear phase error can be calibrated by averaging several CSI phase measurements captured at each channel of certain receiver. To correct the second part of linear phase error, we search for an optimum phase offset that minimize the difference between power delay profile derived from all the available channels under the same multipath environment.

**Multi-Channel Splice**

This step is used to splice together all the available channels at 5GHz band to derive a fine-grained power delay profile. Because of the regulation imposed by different countries, the available channels are unequally and non-contiguous spread across 5GHz band. In particular, the total 24 available channels of 5GHz band on the Intel 5300 NICs are divided into three parts: from channel 36 to 64 (5.17GHz to 5.33GHz), from channel 100 to 140 (5.49GHz to 5.71GHz) and from channel 149 to 165 (5.735GHz to 5.835GHz). The non-available channels are disabled by vendors in compliance with the local regulation. Thus, simply adopting IFFT to transform spliced CSI measurements from available channels to power delay profile is not possible since IFFT only applies to uniformly-spaced frequency measurements.

In our system, we utilize inverse Non-uniform Discrete Fourier Transform(NDFT) which can be applied to non-uniformly spaced channels. To derive fine-grained power delay profile, we denote the CSI measurements from all available channels at 5GHz band as:

$$c_o = [\frac{c_{1,o}}{-}, ..., \frac{c_{n,o}}{-}, ...], \quad (1)$$

where $n$ and $o$ denotes the $nth$ channel at the $oth$ receiver. Given the sampled channel response CSI, the power delay profile $g$ at given channel can be derived using IFFT:

$$g_n = \sum_{l=1}^{L} a_l \delta(t - t_l), \quad (2)$$

where $l$ denotes the sequence number of total L multipath, $a_l$ and $t_l$ are the amplitude and signal propagation time delay of $lth$ path, $\delta(t)$ is the Dirac delta function.

Then, we can formulate the inverse NDFT problem as following:

$$\min_{g} ||c_o - Fg||^2, \quad (3)$$

where $g$ represents the power delay profile we are trying to find and $F$ is Fourier matrix. The goal is to search for an optimum solution of power delay profile that can minimize the difference between the Fourier Transformation of $g$ and spliced CSI measurements from all available channels.

The searching for optimum power delay profile has non-linear and no-closed form solution. Furthermore, the direct search can yield large number of possible results due to large bandwidth of 5GHz channels. In order to filter out the undesirable solutions, we need to include certain constraints to reduce the search space. Previous work utilizes the observation that within indoor environments where only few multipath would dominate the signal propagation [37]. Such constraint works well when the multipath propagation is relative simple (e.g., only a single user inside the room). But it can suffer from performance degradation when there are multiple users within the same environment which creates far more complicated signal propagation.

To overcome such problem, we leverage the layout and distance information between each transmitter and receiver pair. Assuming the signal propagation from the transmitter to one or more receivers has line-of-sight in the system setup, such LoS propagations will dominate the power in the received signal (i.e., the LOS path has the largest power). Therefore, among all the possible solutions of $g$, our system favors the one that has larger power at the LoS propagation. We utilize the proximal gradient method to solve our convex optimization problem [11]. After performing inverse NDFT, we are able to derive a fine-grained power delay profile with the improved time resolution at 5*ns*.

**Reflection Profile Segmentation**

In this step, we separate the signal reflection of each individual user based on the derived fine-grained power delay profile at multiple WiFi links as shown in Figure 3. As the derived power delay profile contains signal reflections from both the static environments and multiple users, we first perform static environment subtraction, where the profile that include multi-user activities is subtracted by the profile under static environment. Note that the profile of static environment can be collected when there is no human present (i.e., the power delay profile under static environment remain constant). After static environment subtraction, we obtain the profile solely containing the signal reflections

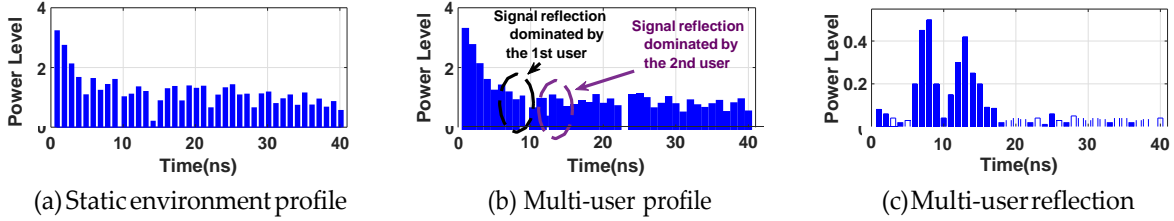

(a) Static environment profile   (b) Multi-user profile   (c) Multi-user reflection

**Figure 3: Illustration of static environment subtraction.**

from multi-user activities, which is referred as multi-user signal reflection profile.

As we can see from Figure 3(c), after static environment subtraction, we are able to observe that there are two major signal reflection components in the multi-user reflection profile. Next, we segment the multi-user signal reflection profile into multiple single user reflection profile with each one representing the signal reflection from one user. By detecting the number of major signal reflection components, we can determine how many users are in the same environment and further segment each user's reflection profile. This is done by using a moving window based approach.

In particular, we accumulate the power differential between adjacent time points within each sliding window. Then we compare the accumulated value to an empirical threshold to determine the duration for each individual profile. In our experiments, the threshold is set to be 0.6 times the standard deviation of the accumulated differential across multi-user reflection profile. After identifying the duration of the desired profile, we are able to determine the number of users and obtain single user reflection profile. We repeat this process over the multi-user reflection profiles derived from all available transmission links.

It is worth noting that, when multiple users are at the same distance with respect to a transmission link, the signal reflection from these users will overlap (i.e., with the similar propagation time delay) in the multi-user reflection profile at that transmission link. Thus simply utilizing power delay profile derived from a single link could not distinguish multiple users under that scenario. Here, we propose to use multiple transmission links (e.g., 3) to overcome this problem. Due to geometric relation between three transmission links, one or more transmission links could capture signal reflection from multiple users without overlapping. Therefore, we can determine the number of users based on these transmission links that do not experience severe overlap.

**Localization and Tracking**

*Close Proximity Separation.* Even after reflection profile segmentation process, it is still possible the segmented reflection profile contains signal reflection from more than one single user, when two or more users are in close proximity (i.e., less than 0.3m). It is difficult to further separate the reflected signal component dominated by individual due to bandwidth limitation of commodity WiFi. Thus it is necessary to address such issue since it is quite common when tracking multiple users within the confined space.

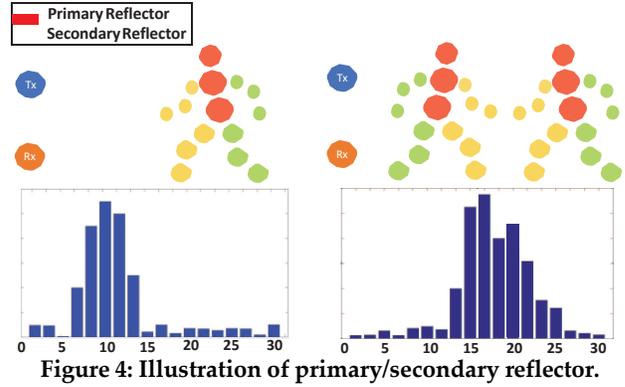

**Figure 4: Illustration of primary/secondary reflector.**

Next, we present the insights that can be utilized to solve the problem of close proximity separation. We know that the received signal can be represented by the sum of multiple components travelling through different paths with varies ToF (time-of-flight). Left side of Figure 4 represents the profile of a single user reflection after static environment subtraction. We can observe that the profile includes reflection path with stronger power which most likely come from user's upper body, and reflection path with weaker power which could be from different limbs of that user. This leads to our first insight: single user body involves different parts that reflect RF signal through different paths. So instead of considering each user as single reflector which contains only dominant ToF path in the previous work, we can further model individual user into combination of primary reflector and secondary reflector.

Furthermore, as users are located within close proximity, the reflection path from different users will be partially combined together due to limited commodity WiFi time/distance resolution. Right side of Figure 4 shows the power delay profile of two users with less than 0.3m distance in between. We can still observe the strong power come from the primary reflector of each user in the multi-user profile. Meanwhile, we can also observe that previous weaker power of secondary reflector path increases due to the inclusion of another user. This leads to our second insight: the increase in the number of users in close proximity would result in a non-negligible increase of the reflection power within that range.

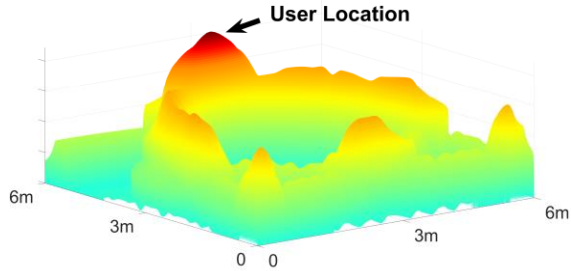

Figure 5: Example of multiple link overlay.

By utilizing those two insights, we can further formulate close proximity separation problem as following:

$$\min_{g_s} ||g_s - R(i,j)||^2, \quad (4)$$

where $R(i,j)$ represents the combined power delay profile for $i$ number of users with $j$ distance in between and $g_s$ represents the mulit-user profile after static environment subtraction. Such power delay profile is acquired through empirical data. The goal is to search for an optimum solution that can minimize the difference between the $g_s$ and combined multi-user profile.

Obviously, such function is non-linear and no closed-form solution exists. Furthermore, it is computation heavy due to the large search space of possible results. In order to solve the problem, we add a constraint which is the derived individual reflection profile should satisfy the geometric relationship between different transmission links. By adding such constraint, we are able to further reduce the large search space and compute the result efficiently.

*Multi-Link Overlay.* Given the single user reflection profile after close proximity separation, we localize each user using multi-link overlay. First, we map all the timed delay power level from the single user reflection profile to the corresponding round trip distance. This will result in a heatmap where strength of the reflection represents the likelihood of the user's location. Then, by overlaying the heatmap derived from multiple transmission links (i.e., 3), we are able to pinpoint each user. Figure 5 shows the overlay results using 3 transmission links where the x and y axis correspond to the localization environment. We can observe that after multi-link overlay, the peak with red color shows the user's location which represents the converge of strong reflector from the individual reflection profile at multiple links. By repeating such process over all the individual profile obtained from reflection profile segmentation process, we are able to localize each user.

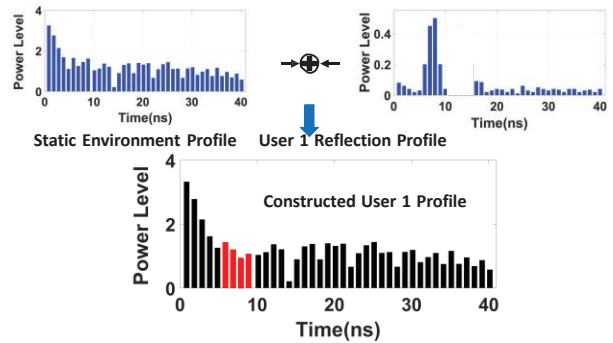

Figure 6: Illustration of individual profile construction.

### Activity Recognition

*Individual Profile Construction.* We perform individual profile construction to obtain signal reflection profile dominated by a single user as if only a single user has performed activity in the environment. To construct individual user reflection profile, we combine the segmented profile from previous step (i.e., single user reflection profile) with the profile of static environment. By doing this, we are able to construct the individual profile containing both signal reflection from target user activity and the static environment but without interference from other user activities. For example, individual profile on the bottom of Figure 6 is constructed by adding static environment profile from top left to the single user reflection profile of the first user from top right.

As demonstrated in Figure 6, after individual profile construction, the profile component in red color represents the signal that is mainly affected by 1st user. Then the profile component in black color is mainly affected by static environment. It is easy to observe from Figure 6, after individual profile construction process, the signal reflection mainly affected by 1st user is preserved whereas the signal reflection most likely affected by the other user is mitigated. The reconstructed individual profile enable us to extract environment-independent features for multi-user activity recognition.

*Doppler Shift based Feature Extraction.* Because of the rich multipath propagation within indoor environment, the constructed individual profile contains signal components reflected by both target motions and the surrounding objects and environment. It is difficult to separate target user's movement from static environments using raw signal. To overcome such problem, we exploit the fact that Doppler shift represents frequency change information of the movement, which wouldn't be affected by signal reflection from surrounding environment. Here, we propose to extract Doppler shift by utilizing short-term Fourier transform (STFT) and compute the spectrogram which is the time-frequency representation of the given frequency response. Specifically, we apply STFT to the individual profile with a Gaussian window with length shorter than 0.1s where we assume Doppler shift is constant within such window.

Figure 7 shows the spectrogram of a user performing lunge exercise towards the transmission link then turning away from it. We can observe that when user's body move towards the receiver, there is an increase follow by decrease positive Doppler frequency shift which indicates acceleration and deceleration motion. Then during the time period that user retracts his/her body away from the receiver, we can observe the similar Doppler frequency shift trend in the spectrogram where the frequency shift is negative.

Next, we extract the energy based frequency contour of derived spectrogram from previous step. To achieve that, we first normalize the energy level of given spectrogram into the same scale (i.e., from 0 to 1). Then we choose a predefined band (power level between 0.90 to 0.95) and combine the centroid frequencies at this band together resulting in two frequency contours (both positive and negative). Such contours represents the strongest signal reflection component caused by user motion as shown in Figure 7. Then the extracted energy-based frequency contour will be used as feature for similarity comparison.

*Similarity Comparison.* As one user may perform activity with different speeds and multiple receivers could be used for activity recognition, we utilize Muti-Dimensional Dynamic Time Warping (MD-DTW) [45] to align the extracted feature to the ones in user enrolled profile. MD-DTW allows us to overcome the problem of pace variety and provides a robust metric for measuring the similarity. In particular, the similarity is quantified by the Euclidean distance of the optimal warping path between the contour and the activity profile. During activity recognition, we extract energy-based frequency contour as feature and use Muti-Dimensional Dynamic Time Warping to calculate the similarity between the feature extracted under training and testing instances. The one with the highest and sufficient similarity (i.e., > 0.73) in the activity profile is then identified as the recognized activity. The one with insufficient similarity (i.e., < 0.73) to existing activity is identified as unknown activity.

## 3 PERFORMANCE EVALUATION
### Experiment Setup

We conduct experiments with four laptops (one transmitter, three receiver) with the default transmitter receiver setup shown in Figure 8. All laptops run Ubuntu 12.04 LTS and are equipped with the WiFi NICs of Intel 5300 for extracting CSI measurements [9]. The transmitter hops through all available 20MHz WiFi channels at 5GHz bands in an 802.11n network. There are total 24 available channels enabled by the Intel 5300 card. They fall into three non-contiguous segments. The first segment is from 5.18GHz to 5.32 GHz (i.e., the channels from 36 to 64), whereas the second segment is from 5.5GHz to 5.70GHz (i.e., the channels from 100 to 140). The third one is from 5.73GHz to 5.83GHz (i.e., the channels from 149 to 165). The channel hopping delay is set as 0.2ms. As the coherence time in typical indoor environment is about several hundreds milliseconds [8], we can collect packets across channels within coherence time as well as obtain multiple packets at each channel within coherence time. For each packet, we extract CSI for 30 subcarriers, which are equally distributed in a 20MHz channel.

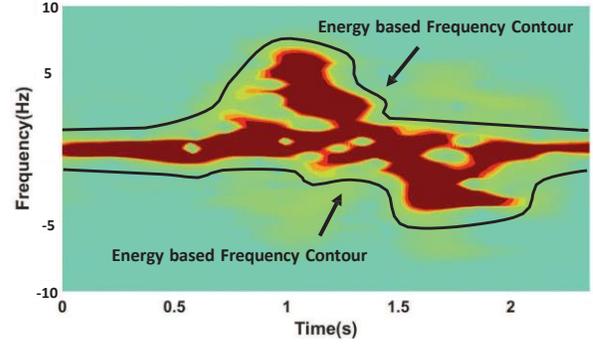

**Figure 7: Spectrogram of lunge with energy based frequency contour.**

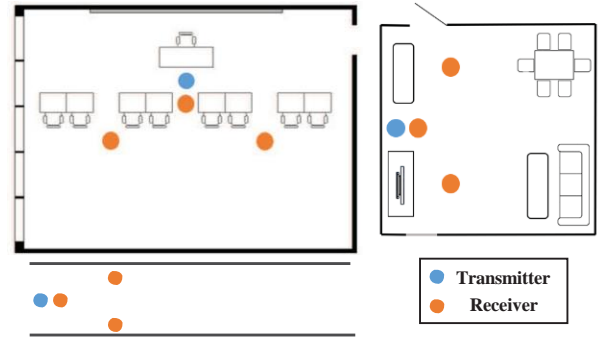

**Figure 8: Illustration of experiment setup.**

To evaluate the performance of MultiTrack, we conduct separate experiments to test the localization and activity recognition components. For localization and tracking component, we conduct experiments in 3 indoor environments: a 25ft by 30ft classroom, a 15ft by 20ft living room and a narrow corridor. Figure 8 shows the deployment of devices in different environments. In total, 5 volunteers (3 males and 2 females) participate in the experiment. We obtain the ground truth through camera-based tracking techniques. In particular, a video camera is installed to capture the tracking process. Each volunteer is instructed to wear different color hats as markers for easier identification purpose. Then, we convert the pixel location of the marker into real world 2D location which served as ground truth.

For activity recognition component, the experiment is conducted in the living room environment. We evaluate the

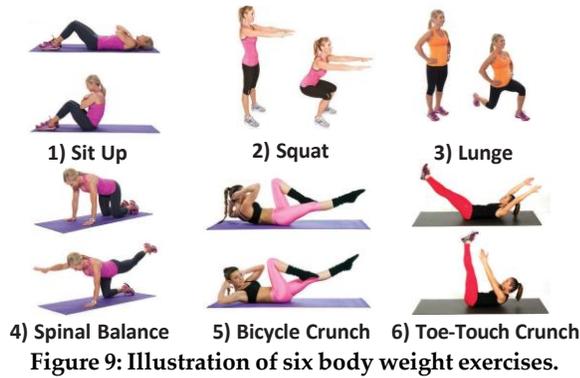
**Figure 9: Illustration of six body weight exercises.**

performance of our system with six commonly used body weight exercises including: sit up, squat, lunge, spinal balance, bicycle crunch and toe-touch crunch, as shown in Figure 9. As a benchmark, we ask each volunteer to perform one activity fifty times alone, and ten instances of each activity are used to build the activity profile. To test the multi-user compatibility of our system, we experiment with the scenario where three users are performing different exercises simultaneously. Furthermore, to evaluate the robustness of our system to unknown activities, we conduct the experiment that two users are performing pre-defined exercises and the third user is performing unknown activity simultaneously.

The detail about the data for activity recognition are as following: for single user case (i.e., only one user is performing activity), each volunteer (5 in total) is asked to perform each of six types of activities 50 times (300 instances for each type of activity). Then, for each type of activity, we randomly select 10 instances as training data to build the activity profile (i.e., the profile is not tied to each user), which is used for both single user and multi-user activity recognition. The rest of the 290 instances for each type of activity are used as the testing data for single user case recognition.

For multi-user case, there are two experimental setups. The first setup includes three users performing activities simultaneously: two users randomly perform six types of activities while the 3rd user preforms unknown activities (i.e., serving as background interference) for 50 times (150 instances for each type of activity). The second setup includes two users perform activities simultaneously without unknown background activity for 50 times (100 instances for each type of activity). In total, we have an averaged 250 instances for each type of activity under multi-user case.

### Multi-User Tracking Performance

We first show the results of tracking a single user under different environments for comparison. Figure 10 shows the localization error CDF of our system at different places under single user cases. We can observe that MultiTrack achieves localization errors of 0.39m, 0.57m and 0.65m over $80^{th}$ percentile. When compare with state-of-the-art device-free commodity WiFi based single user tracking systems (i.e., Widar 2.0 [29], IndoTrack [20], LiFS [42] and WiDeo [14] with median accuracy as 0.75m, 0.48m, 0.7m and 0.7m), our system achieves better or comparable performance with a median localization accuracy at around 0.5m. This is because our system utilizes a much larger channel bandwidth at 5GHz, which provides fine-grained power delay profiles to characterize user's location.

Next, we evaluate the localization performance in multi-user scenarios. The experiment is conducted in classroom environment as it allows multiple users to walk simultaneously. The CDF of localization error is plotted in Figure 11. We observe that the median localization error for 2, 3 and 4 users are 0.46m, 0.55m and 0.81m respectively. This demonstrates our system has the ability to track multiple users simultaneously with decimeter localization accuracy. Moreover, under multi-user case, our system performance is also comparable to the results under single user case for both the state-of-art systems and our own system.

### Multi-User Tracking with Close Proximity

Next, we study how our system performs when two users are in close proximity in the living room environment. Specifically, the distance between two users are 1m, 0.5m and less than 0.5m. Figure 12 shows the performance of our system under these scenarios. In particular, the median error of our system is 0.25m, 0.3m and 0.35m when the distance between the users are 1m, 0.5m or less than 0.5m respectively. Such result shows that our system achieves high accuracy even when users are in close proximity. We note that the localization accuracy in living room environment is better than that of the classroom. This is because the living room is much smaller but with the same number of wireless links. Thus, a higher density of wireless links in the environment could also help to improve the tracking accuracy.

### Multi-User Tracking Under NLOS

In this study, we investigate the impact of NLOS to tracking accuracy by placing the WiFi devices in two connected rooms with line-of-sight blocked. The experiment is also conducted in classroom environment. Figure 13 presents the CDF of localization errors under different number of users. Results show that under NLOS scenarios, system performs slightly worse. Still, NLOS scenario has the median localization error of 0.46m, 0.52m and 0.61m with respect to 1, 2 and 3 users scenarios. This demonstrates that the our system could work under the NLOS scenario which allows us to deploy the proposed system to a wider range of applications.

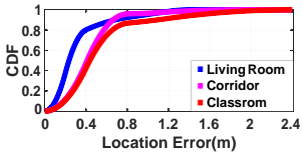
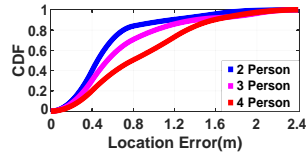
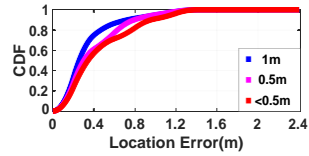
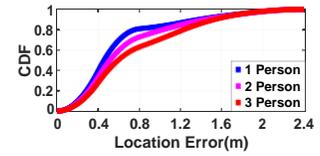

Figure 10: Different environments performance.

Figure 11: Multi-user scenarios performance.

Figure 12: Close proximity scenarios performance.

Figure 13: NLOS scenarios performance.

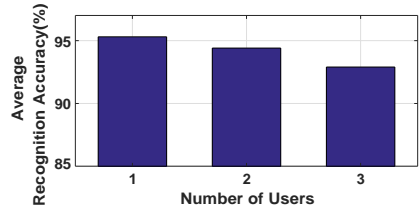
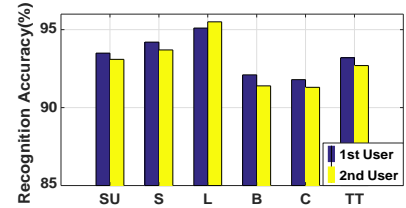

Figure 14: Single user activity recognition performance.

Figure 15: Recognition accuracy under multi-user scenario.

Figure 16: Recognition accuracy when other user perform unknown activity.

### Multi-User Activity Recognition

We first present the results of activity recognition when there is only a single user performing activity in the environment as a benchmark. Figure 14 shows the confusion matrix of activity recognition accuracy under single user scenario. We observe that our system achieves an overall recognition accuracy of 95% with the standard deviation at about 1.2%. By comparing the details across different activities, we find that the recognition accuracy are comparable. Moreover, activity like sit up, squat and lunge have higher recognition accuracy, whereas the bicycle crunch and spinal balance have lower accuracy. In particular, the lunge achieves 97% accuracy compared to 94% recognition accuracy of spinal balance. This is possibly due to the relative larger body motion involvement in exercise like lunge. Consequently, more details of the target motion could be captured by CSI measurement.

Figure 15 illustrates the average recognition accuracy across different activities when multiple user scenarios (i.e., 2 or 3 users) are performing the activities at the same time. We can observe that our system can maintain high accuracy even when there are three users performing different exercises simultaneously. Specifically, the average recognition accuracy is over 94% for two users whereas it is 95.3% for single user scenario. Furthermore, the accuracy only drop by 2% when three users are performing different exercise simultaneously compared to single user case. The above results show our system can recognize activities of multiple users preforming them simultaneously with high accuracy.

### Impact of Unknown Activities

We next test the resilience of our activity recognition system to unknown activities. We ask the third user to perform unknown activities while two other users are performing different per-defined exercises at the same time. Figure 16 presents the recognition accuracy of each exercise for each individual user. We find that the recognition accuracy of our system is comparable to the single user scenario. Moreover, by comparing Figure 16 and Figure 14, we observe that the system performance does not have obvious degradation even under the interference of the unknown activity performed by the third user. This study demonstrates our system can recognize the activities of target users when there are unknown background activities.

## 4 RELATED WORK

In general, the approach for indoor tracking and activity recognition can be divided into following categories based on its underlying technique: camera and visible light based, wearable device based and RF signal based.

**Camera and visible light based.** Much work has been done to enable indoor tracking using dedicated cameras [31, 33]. With the advancement of imaging technology, recent work like Eaglesense [47] and OpenPTrack [24] can track multiple people within the same environment. Though these approaches can achieve high accuracy, it raises great concern of user privacy and only work under LoS scenarios. Recently, due to fast development of visual light communications, visible light based approach has attracted lots of research interests [18, 53]. Such work still can not work under NLoS scenario and require specialized light source which incur non-negligible cost and installation overhead.

**Wearable device based.** With the advancement of wearable devices, many work have been proposed to solve tracking and activity recognition problem. For example, Ashbrook

*et al.* [4] proposed work the predict user future movements based on the individual GPS trajectories. TouchRing [36] focus on subtle finger movements detection with ring shaped printed electrodes wore by users. Baudisch *et al.* [5] invented an imaginary ball game using accelerometers attached to the players hands and belt. Though effective, these methods all require users to wear physical sensors.

**RF signal based.** Device-free indoor tracking utilizing COTS hardware has been an active research field in recent years. Kjærgaard *et al.* proposed several work [16, 26] on WiFi positioning and flock detection. System like LiFS [42] leverage signal phase characteristic in and out of the Fresnels Zone to achieve tracking but requires specific and dense deployment of WiFi devices. Li *et al.* [19, 20] proposed several work by incorporating DFS with AoA to achieve indoor tracking. Qian *et al.* [28, 29] developed a serious of systems that can jointly estimate AoA, DFS and ToF which enables decimeter tracking. These systems, however, can not work under multi-user scenarios.

Besides indoor tracking and localization, many research have been dedicated to achieve activity recognition using commodity WiFi as well. Wall++ [52] proposed by Zhang *et al.* can achieve context-aware sensing by capturing airborne electromagnetic noises. Systems like WiFall [46] E-eyes [45] and CARM [44] are able to recognize large scale motions (i.e., falling, walking, etc.) Moreover, several work has been proposed to enable vital sign monitoring related applications [21, 41]. System proposed by Wang *et al.* [40] can recognize the words user spoke by tracking the mouth movement. WiFinger [35] is capable of tracking small scale finger motions. Those systems however all require localization-specific training and cannot support multi-user scenario.

Much research has been dedicated to solve environment-dependent and multi-user compatible problem using commodity WiFi. Systems like CrossSense [50] and EI [13] adopt deep learning techniques to achieve tracking accuracy improvement comparing to existing systems. However, they all require large number of training samples and constant update when environment changes. Virmani *et al.* [39] proposed WiAG that utilizes profile transition function to estimate user's new orientation and position once change happens. Meanwhile WiDance [30] achieves environment-independent sensing by extracting motion-induced Doppler shift to enable dancing move direction recognition. The systems mentioned above though can be effective, all lack multi-user support. System like WiMU proposed by Venkatnarayan *et al.* [38] can achieve multi-user gesture recognition by matching the generated virtual samples of desired gesture combination to the collected samples. The proposed system partially solved multi-user compatible problem. But it can only work when system has pre-knowledge of all possible activities. It cannot work when non-target users perform unknown activities in the background.

## 5 DISCUSSION

**Multi-User Support Capability.** The underlying idea of our system is to separate signal reflection for each individual user. As the WiFi bandwidth at 5GHz provides a distance resolution at around 0.3 meters when separating signal reflections, our system thus is limited when multiple users are too close to each other (e.g., less than 0.3 meters). We have tested our system in a typical indoor environment (i.e., a 15 ft by 20 ft living room) with up to four users. This is a typical use case as the system can support a reasonable number of users that living together in indoor environments (e.g., smart-home environment). And such a case provides sufficient physical distance separation for our system to work. On the other hand, with very crowded environments, such as classroom, a train station or an airport, our system is less applicable due to extremely small distance separations between users.

**Machine Learning Techniques.** The current activity recognition process using Euclidean distance based method to calculate the similarity between testing samples and user enrolled profile is relatively simple. A more sophisticated machine learning based classification method (i.e., convolutional neural network) could be used to increase recognition accuracy and system resilient to noises. We would like to include this part as our future work to further improve system accuracy and robustness.

## 6 CONCLUSION

This paper presents MultiTrack, which is capable of tracking multiple users and recognizing activities of multiple users perform them simultaneously. The proposed system doesn't require user to carry or wear any dedicated sensors and can reuse existing commodity WiFi devices. The insight is that the signal reflection from each individual user in multi-user scenarios can be extracted with the help of multiple WiFi links and the large bandwidth at 5GHz. By analyzing the extracted signal reflection of each user, MultiTrack achieves multi-user tracking and activity recognition. Experimental results under different multipath environments show that MultiTrack can achieve decimeter localization accuracy when 4 users are walking simultaneously. Meanwhile, MultiTrack can achieve activity recognition accuracy over 92% when multiple users are performing activities concurrently with background interference from unknown activities. In addition, we show that our system can work under NLOS scenario and provide considerable localization accuracy even when users are in close proximity.

## ACKNOWLEDGMENTS

We thank the anonymous reviewers for their insightful feedbacks. This work was partially supported by the NSF Grants CNS-1505175, CNS-1514238 and DGE-1565215.